# Supporting 'Good Habits' through User-Led Design of Food Safety Applications – Findings from a Survey of Red Meat Consumers

**Adeola Bamgboje-Ayodele**
School of Engineering and ICT
University of Tasmania
Hobart, Australia
Email: Adeola.Bamgboje@utas.edu.au

**Leonie Ellis**
School of Engineering and ICT
University of Tasmania
Hobart, Australia
Email: Leonie.Ellis@utas.edu.au

**Paul Turner**
School of Engineering and ICT
University of Tasmania
Hobart, Australia
Email: Paul.Turner@utas.edu.au

## Abstract

Mitigating consumer health risks and reducing food wastage has stimulated research into mechanisms for improving consumers' food safety knowledge and food management practice. Many studies report success, but differences in methodology and in the type and range of foods and consumers involved has made comparison and transferability of results challenging. While most studies advocate for the importance of information in consumer education, few provide detailed insight into what 'good' information means. Determining appropriate content, formats, and methods of delivery for different types of consumers as well as evaluating how different choices impact on consumers' food safety knowledge and behaviour remains unclear. Within a larger research project on enhancing provenance, stability and traceability of red meat value chains, this paper presents findings from a survey of Australian red meat consumers (n=217). It identifies consumers' food safety issues and reveals information and communication preferences that may support good safety habits with food.

**Keywords**

Food Safety, Safe Food Handling, Consumer Behaviour, Usability, Information Technologies.

## 1 INTRODUCTION

Research into differences amongst consumers in their understanding of food safety and practice of food management recognises the role of socio-economic, cultural and educational factors in shaping individuals knowledge, attitudes and behaviours. Beyond consumers themselves, it has also been recognised that some of the difficulties and confusion about food safety and about appropriate food management practices reflect problems with existing regulations and industry practices. For example, *'Made in' Australia* on a food product actually only means that "the product was made/packaged in Australia and that at least 50% of the cost to produce it was incurred in Australia" in other words, the food product may not necessarily contain any Australian ingredients yet be labelled made in Australia (ACCC 2015).

Aligned to concerns about consumers understanding of food safety are broader issues related to food wastage all along contemporary agri-business food supply chain including by consumers in their own homes (Gustavsson et al. 2011; Parfitt et al. 2010). Varying degrees of consumer knowledge about food labels, food storage, portion control and home food management as a whole directly influence consumers' propensity to waste food (Monier 2011). More specifically, in times of uncertainty such as a food recall situation, consumers make decisions based on their understanding of the available information on food products. Therefore, limited, incomprehensible or poorly structured information has the potential to increase food wastage in normal day-to-day activities and especially during food recall incidents (Lyndhurst 2008; Monier 2011).

In terms of mitigating consumer health risks from food poisoning, while regulations primarily focus on food production, processing and retail, it is now increasingly recognised that domestic mismanagement of food products from the point of purchase through to actual consumption is a growing source of concern. According to the New South Wales Food Authority, foods can become contaminated while in the possession of consumers in a number of ways including; not cooking food thoroughly, not storing food appropriately, poor hand hygiene, eating food after a 'use-by' date and cross contamination between foods (NSWFA 2014). Out of the reported cases of food borne outbreaks in 2010, 2146 persons were affected, including 157 hospitalizations and 15 deaths (OzFoodNet 2012). Apart from commercially prepared food, the consumer home has the highest percentage of food poisoning outbreaks in Australia (OzFoodNet 2012).

In this context, and despite considerable efforts with regards to public food safety information campaigns, many consumers remain inadequately informed about food safety and continue to engage in unsafe food handling practices. Unsurprisingly, this has led to numerous research studies focused on addressing consumers' impaired decision-making due to lack of food safety knowledge (Bondarianzadeh et al. 2011; Ergönül 2013; Losasso et al. 2012; Mateus et al. 2014; Nesbitt et al. 2014; Ovca and Jevšnik 2009; Shim et al. 2011; Stenger et al. 2014; Taché and Carpentier 2014). Despite differences in methods and in the type and range of foods and consumers studied, most of these studies advocate for improved education and awareness amongst consumers to address the challenges faced. Unfortunately, few provide detailed insight into what 'good' or appropriate information should be used to ensure effective and sustainable consumer education on food safety. Determining appropriate content, formats, and methods of delivery for different types of consumers as well as evaluating how different choices impact on consumers' food safety knowledge and behaviour continues to be a focus of debate.

This research–in-progress paper, is part of a large Australian Research Council Industrial Transformation Research Hub (ARC-ITRH: Pathways to Market) focused on enhancing provenance, stability and traceability of Australian red meat value chains into international export markets (Mirowski et al. 2014). The paper presents findings from a survey of Australian red meat consumers (n=217). It identifies consumers' food safety issues and reveals information and communication preferences that may support good safety habits with food. Following an extensive review of the literature on topics of relevance to this investigation in previous publications (Bamgboje-Ayodele et al. 2014; Bamgboje-Ayodele et al. 2015), this paper progresses by presenting a summary of the review, the method of research and findings of the survey. Next, it discusses implications for the developing over-arching methodology and innovative approaches for educating food consumers before concluding the paper.

## 2 FOOD SAFETY, TECHNOLOGY AND CONSUMERS

In advocating for improved education and awareness amongst consumers on food safety, a number of researchers have explored a role for information systems. A review of recent research literature in this area identifies a number of key areas where technology has been deployed such as; labels, information sheets, barcodes, 2D barcodes and RFID tags. These technologies have been researched in relation to nutritional information but a number of gaps remain including detail on preferences of consumers in the use of QR codes and/or NFC tags with mobile devices (Chrysochou et al. 2009). There is however some evidence (Voordouw et al. 2011) suggesting that some consumer's preference for the use of labels and information sheets is linked to their educational and technological backgrounds, although it is unclear whether these insights are culturally constrained or can be applied to Australian consumers. Based on the results of the study conducted by Reid et al. (2001), it appears that the application of food related lifestyle consumer categorization is also an important dimension. Similarly, Van Rijswijk and Frewer (2012) have emphasized the importance of consumer categorization as a pre-requisite for determining the influence of any information delivery platform on different types of consumers.

Research work in the logistics domain has investigated the impact of traceability through technologies such as laser technologies, DNA based technologies and e-paper tags, on consumers' perception (Chrysochou et al. 2009). Other researchers have also explored consumer perceptions about the use of internal tag technology involving insertion of 'internal tags' into food and natural health products (Lilavanichakul and Boecker 2013). However, to date, these approaches have been primarily focused on pre-programmed information focused on information about nutrition, producer and country of origin. Regardless of the quality and safety of any food product, if it is handled inappropriately post-purchase, it may well lead to health related issues. Therefore, it remains unclear how these emerging technologies might be deployed and how they would influence consumers' safe handling of a food product post-purchase.

To bridge the gap between technology and consumers, one important area of focus is usability evaluation. While research work on usability testing deals with ensuring that products meet the actual needs of the intended audience – in this case the consumers (Rogers et al. 2011), there is limited evidence of these approaches being used for food consumers. Indeed, the focus of many of these studies has largely been on the adoption and expected use of these technologies using a 'one-size fits all' approach. There have been fewer studies trying to focus on design and implementation in a way that facilitates intended use by different types of consumers.

Numerous studies have offered the same recommendation towards addressing consumers' inappropriate food handling practices due to lack of food safety knowledge. They have argued that education and awareness will address the problem (Bondarianzadeh et al. 2011; Ergönül 2013; Losasso et al. 2012; Mateus et al. 2014; Nesbitt et al. 2014; Ovca and Jevšnik 2009; Shim et al. 2011; Stenger et al. 2014; Taché and Carpentier 2014). However, none of these studies clearly elaborated on practical measures that are helpful in establishing such education and awareness programs and its complexities. In addition, they have not been able to provide significant evidence to suggest that consumers indeed understand the information and can retain the knowledge acquired through these education and awareness programs over time.

While it has been argued that the amount and accuracy of consumers' knowledge does not necessarily translate to corresponding behaviour (Aizaki et al. 2011; Al-Sakkaf 2012), many research advocates have argued for the use of *theory-driven interventions* (social cognition theories such as the health belief model (Rimal 2000), health action process approach (Bearth et al. 2014) and theory of planned behaviour (Sainsbury et al. 2013)) in designing food management and safety interventions. However, there is evidence to suggest that theory-driven interventions do not necessarily improve consumer behaviour. For example; to investigate the influence of a food hygiene intervention on human behaviour, Mullan and Wong (2010) and Phillip and Anita (2010) utilized the theory of planned behaviour to design the intervention. However, their findings indicate that the intervention did not improve the behaviour of the participants.

On the other hand, recent research into the use of *habit-driven interventions* have provided evidence to suggest that interventions that provide cues to actions and reminders that build food safety and management habits result in changes in behaviour. For example; Rompotis et al. (2014) successfully used habit formation to improve fruit and vegetable consumption behaviour through the use of SMS and email as information delivery channels. Also, Mullan et al. (2014) successfully used habit strength to improve food safety behaviour through the use of poster and email as information delivery channels. However, the participant group for their study was limited to undergraduate students in a university. Therefore, one can argue that a more thorough understanding may be gained if the sample covers a broader population. Nonetheless, it appears habit-driven interventions are more promising in facilitating an improvement in safe food handling practices.

Drawing on these habit-driven studies, one common theme is the use of information delivery tools and information modalities which help to provide cues to action and reminders. However, Mullan et al. (2014) did not investigate the influence or differences between the use of textual and visual information cues (poster) and text-based cues only (email). For Rompotis et al. (2014), they found that there were no differences between SMS (text-based) and email (text-based) delivery channels. However, it remains unclear how visual, verbal and integrated information forms and other IT delivery channels would influence consumer safe food handling practices (post-purchase) as studies in the Information Systems discipline suggest that textual (Blanco et al. 2010), visual (Ha and Lennon 2010; Lin et al. 2012) and verbal (Kim and Lennon 2008) modes of information influence consumer pre-purchase behaviour. Hence, it will be interesting to investigate the influence of information modalities and information delivery channels through habit-driven interventions on consumer safe food handling practices (post-purchase). Thus, this leads to the question; how does the use of information technology in habit-driven interventions influence consumer safe food handling practices? While it is acknowledged that consumers' knowledge does not necessarily translate to corresponding behaviour, it can however be argued that ensuring knowledge retention on safe food handling practices is one step closer to achieving corresponding behaviour. *Therefore, this research investigates the impact of a usability evaluation informed design approach on consumers' retention of knowledge on safe food handling practices.*

More importantly, as this brief review of multi-disciplinary research on the challenge of food safety, technology adoption & use and consumers, highlight numerous gaps remain in our knowledge. It is unclear how consumers differentially make sense of information acquired through the use of technology and how they respond to it based on their individual socio-economic, educational and cultural contexts. In this context, this research paper aims to contribute to mechanisms for enhancing consumer-focused technology-based information delivery mechanisms and contemporary understanding of how the attributes of different consumers influence their responses and decisions around food safety post-purchase. The next section describes the methodology deployed for the conduct of the survey.

## 3　METHODOLOGY

This research has been approved by the Tasmanian Social Sciences Human Research Ethics (H0014658). The approach utilised a pragmatic research philosophy and deployed a mixed-method design structured in three overlapping phases. Phase 1 (Consumer Understanding) involved the conduct of a nationwide survey to identify problems with the current food handling practices of Australian consumers and their information and communication preferences (both pre-purchase and post-purchase). This survey is primarily focused on the consumption of red meat due to the enormous risks associated with its handling and management. The outcome of this phase provides insight into the most problematic food handling practice for Australian consumers, so as to validate the existing literature on food poisoning and food wastage *(see section 2)* and to scope the investigation to the most significant challenges. This leads to the selection of three existing apps (text-based, graphics/picture-based and integrated) that most clearly addresses the safe food handling practice being targeted. Secondly, this phase will also provide a broad insight into consumers preferred style of design.

Phase 2 (Design) involves the heuristic evaluation of the three existing apps based on Monkman and Kushniruk (2013) framework to identify problems with the apps from an expert's perspective. Following this, a second usability evaluation from the consumers' perspective, using the apps as a high fidelity prototype in scenario-based focus group sessions, will be conducted. This research activity aims to identify the impact of the three information modalities on consumer understanding and to generate user requirements for an app. The outcome of this phase provides rich insights into consumer requirements for a safe food management app. This leads to the design of a single smartphone application for educating and assisting consumers on the safe food management practice identified from the survey in Phase 1 as the most problematic.

Phase 3 (Development and Implementation) involves the actual development and evaluation of the app designed in Phase 2. After it has been developed, it will be evaluated by conducting a longitudinal experiment, within a 10-week period, using an experimental group (participants involved with the design in Phase 2) and a control group (new set of participants). The aim of this research activity is to achieve two things: evaluate the usability of the design; and to evaluate the impact of the design on the retention of knowledge on safe food handling practices over time.

To date, Phase 1 (Consumer Understanding) has been completed and will be discussed in more detail below. While the survey method is typically associated with the positivist paradigm, it has been identified as a method useful in the pragmatism paradigm. Consistent with this, Charmaz (2006) has argued that the choice of method should be based on the research objectives and not on pre-conceived ideas about what seems more appropriate according to the philosophical paradigm. Hence, survey has been adopted as a data collection tool for the Phase 1 of this study. Within this phase, the best worst scaling survey approach has been adopted. Best Worst Scaling (BWS) also known as Maximum Difference Scaling (Maxdiff) is a measurement or scaling technique originally developed by Jordan Louviere and others (Finn and Louviere 1992; Louviere and Woodworth 1991). This technique has recently gained grounds within the food and health economics literature (Erdem et al. 2012; Louviere and Flynn 2010; Lusk and Briggeman 2009) but not so much in the information systems discipline. The aim of this technique is to elicit the relative weight or importance that participants allocate to attributes in large sets (Erdem et al. 2012).

The BWS technique has several advantages over other approaches such as the rating based methods (Cohen 2003; Erdem et al. 2012; Lusk and Briggeman 2009; Marley 2009). A rating based technique provides a scale of 1 to 7, for example, with 1 being least important and 7 being the most important to respondents. One of the disadvantages of the rating based methods is that participants are not forced to make trade-offs between the relative importance of issues (Lusk and Briggeman 2009). Another disadvantage of rating based methods is that different respondents utilize such scales differently as what represents 7 for an individual may represent 5 for another. However, the BWS technique avoids these drawbacks by forcing participants to respond with a choice (Lusk and Briggeman 2009; Marley 2009). Furthermore, this approach is useful when seeking preferences over a large number of attributes. Ranking attributes according to their importance may become difficult as the number of attributes increases. However, the BWS technique breaks this task into a more cognitively manageable size, thus making tasks easier and less prone to anomalous behaviour for the respondents (Erdem et al. 2012).

## 3.1 Participant Selection

The survey was targeted at consumers within a consumer panel database who met the inclusion criteria. In total, 217 of 278 (78%) eligible consumers within the consumer panel cohort responded. The consumer panel utilized is The Online Research Unit which holds ISO 20252 'Market Research Standard' and ISO 26362 – 'Global Access Panels' accreditations and they are also endorsed by the Association of Market and Social Research Organisations (AMSRO) which demonstrates their level of adherence to privacy and ethical standards. This organization was utilized because of a number of reasons such as; their invitation-only policy which increases representation and minimizes self-selection bias; and their primarily offline panel recruitment policy which avoids a sample bias of being over representative of urban areas, heavy online users and young consumers.

In order to ensure the right respondents indeed filled the survey, several measures were taken. Firstly, the consumer panel utilized for the data collection ensured that participants are living in Australia and they are above 18 years of age. Secondly, the participants were quizzed about their dietary intake. This is to ensure that survey respondents are not vegans, vegetarians, or only white meat eaters, as the survey will not be related to them. Thirdly, the participants were also asked about where they shop, as a screening question, in order to ensure that online grocery shoppers were excluded from the survey. The nature of the research context relates to how fresh meat is usually handled from the point of purchase to the home. Online grocery shopping is outside of the scope for which the survey is intended.

## 3.2 Research Instrument

The research instrument was derived from the existing literature and was modified to fit the context of this research. Hyman et al. (2006) have stated that using pre-existing questions provide accurate measure as they are pre-tested before first usage, such that the degree of validity and the quality of the data are likely to be high. Hence, out of the various approaches to content validation discussed by Grover (1997) and Wynd et al. (2003), the content validity of the instrument in this study was addressed by using pre-existing questions and the literature from previous similar studies which was modified to suit the current study.

There are 4 sections of the survey; demography, pre-purchase questions, post-purchase questions and information & communication related questions. The demographic questions contained gender, age group, education, geographical location and household income. The pre-purchase questions provide an understanding of the factors that facilitate food, but more specifically meat purchase of the research participants and were derived from Grunert et al. (2014); Smith and Riethmuller (1999); Williams et al. (2004). The post-purchase questions provide an understanding of the food handling practices of the research participants post-purchase, thus providing a focus for the content of the app that will be developed. This section contained questions on purchase, transportation, storage, thawing practices, cooking and re-heating left overs, kitchen hygiene practices and were derived from Gilbert et al. (2007) and Williams et al. (2004). The information and communication related questions provide insights into the style of the design for the mobile phone application that will be developed (Blanco et al. 2010; De Jonge et al. 2007; Ha and Lennon 2010; Henderson et al. 2011; Jacob et al. 2010; Kuttschreuter et al. 2014; Lin et al. 2012).

Drawing on the pioneering study of the importance of pilot studies by van Teijlingen and Hundley (2001), the product of the modified instrument underwent a pilot study. This was required to test the adequacy of the research instrument and was carried out between February and May 2015. The outcome of the pilot study, led to the final refinement of the survey. On completion of the refinement, the instrument was administered to the respondents via the Internet using the Survey Gizmo website in June 2015. Out of about 278 attempts to complete the survey, 61 responses were incomplete; thus resulting in 217 completed responses. After the data collection, the data was exported to Microsoft Excel 2010 for initial formatting and then imported into IBM SPSS software version 22.0 for better analysis.

The post-purchase and demographic questions were analyzed using descriptive statistics; pre-purchase questions were analyzed using Maxdiff analysis while the information & communication related questions utilized both modes of analysis. Maxdiff data was analyzed using the Hierarchical Bayes Model to compute the Bayesian average. Calculating the Bayesian average used the prior mean $m$ and a constant $C$. $C$ is assigned a value that is proportional to the data set size. The value is larger when the expected variation between data sets (within the larger population) is small. It is smaller, when the data sets are expected to vary substantially from one another.

$$\bar{x} = \frac{C_m + \sum_{i=1}^{n} x_i}{C + n}$$

The result of the analysis is summarized in the next section.

## 4 RESULTS

This section presents the results of the data collected in Phase 1 of the study. Due to space constraints, the results are summarised. As stated in the methodology, these are categorised into the following sub-sections respectively: demographics, pre-purchase related questions, post-purchase related questions (transport, storage, thawing, cooking, reheating leftovers and kitchen hygiene) and information & communication related questions.

### 4.1 Demography

The result shows that 53% of the respondents were females, 46% were males and 1% preferred not to disclose their gender. All the respondents live in Australia. Half of the respondents were aged 50 years and above, 31% were within the age range of 30 to 49 while 19% were between 18 to 29 years. In addition, the participants are located in all the states and regions of Australia except Northern Territory. About 70% of all the respondents were from the southern states within Australia, where climate, culture, eating and shopping habits are likely to be more similar and relevant to beef than those in tropical region. More than 65% of the participants are well-educated. There is a good spread of income across the respondents. 70% of the respondents buy their meats from supermarkets, 29% from butchers and about 1% from delicatessens.

### 4.2 Pre-Purchase Decisions

This section presents the analysis of the two elements that were the focus of the pre-purchase food related questions presented to the research participants using the best-worst scaling technique. The two elements are; attributes and labels considered prior to food purchase.

#### 4.2.1 Pre-Purchase Attribute Preferences

Attributes considered prior to food purchase aims to determine the most important and least important information, which respondents usually consider before the purchase of meat products. The findings reveal that the clear leaders are the use-by/best-before date, country of origin and nutritional/health benefits with Bayesian averages of 8.4, 7.1 and 6.7 respectively. The next group of preferred food attributes were the discounted products, ingredient lists, ethical impact and cooking instructions respectively while the brand and environmental impact information were least preferred.

Therefore, these findings have shown that the most preferred information attributes is the use-by/best-before dates, thus complementing the existing literature (FSAI 2003) on the high level of preference Irish consumers portray towards use-by dates.

#### 4.2.2 Pre-Purchase Label Preferences

Labels considered prior to food purchase aims to determine the most important and least important types of labels, which respondents usually consider before the purchase of meat products. The three types of food related labels in this question are the safety or nutritional label, the environmental sustainability label and the ethical or religious label. The findings reveal that Meat Standards Australia (MSA), National Heart Foundation (NHA) and RSPCA Approved Farming were the most preferred pre-purchase labels with Bayesian averages of 16.1, 15.4 and 12.6 respectively. Ethical labels such as Livestock Welfare Certified Systems and Animal Welfare were preferred next whilst religious labels (Kosher Australia and Halal Australia) were least preferred.

Therefore, these findings have shown that Meat Standards Australia label, provided by Meat and Livestock Australia (MLA), a marketing and research & development body for Australia's red meat and livestock industry, is the most preferred pre-purchase label. This portrays a high level of trust in the MSA system which primarily focuses on meat quality (Polkinghorne et al. 2008; Watson et al. 2008).

## 4.3  Post-purchase Meat Handling

This sub-section presents the analysis of the six elements that were the focus of the post-purchase food handling practice questions presented to the research participants. The six elements are; transport, storage, thawing practices, cooking, reheating leftovers and kitchen hygiene. Due to space limitations, and the relevance of such findings for the Information Systems discipline, only a summary of the findings will be presented in this paper.

The post-purchase data analysis has revealed that the three most problematic food handling practices are cooking, kitchen hygiene and storage. **For storage**, the findings reveal that those (72%) who, had problems with safe packaging of fresh meat for storage in the fridge and those (44%) who do not understand the importance of the location where the meat is placed in the fridge, were within the subset of the 26% of all the respondents, who chose chilling as their preferred mode of storage. Consequently, the weight of this finding is quite low. **For kitchen hygiene**, the findings indicate that only 9% of all the respondents use gloves and at least 80% of these (7.2% of all respondents) do not have the recommended safe kitchen hygiene practices. However, 91% of all the respondents do not use gloves and 55% of these (50% of all respondents) do not wash their hands frequently during fresh meat preparation. Also, it was revealed that 70% of those who do not use gloves (64% of all respondents) do not know how to wash their hands appropriately during fresh meat preparation. Consequently, at least 64% of all the respondents do not have safe kitchen hygiene practice and the weight of this finding is slightly above average. **For cooking**, the findings revealed that 48% of all the respondents undercook their meat when pan-frying a 3.5cm fillet steak. In addition, the findings reveal that 94% of all the respondents utilize diverse ways of evaluating the level of doneness of meat, while it is being cooked, which are against the recommended practice. Out of the 6% who use the recommended method of evaluating how well cooked meat is, at least 50% of these (3% of all respondents) cook their meat to temperature that is lower than the best practice. This results in 97% of all the respondents who do not understand the most appropriate method of evaluating the doneness of meat. Consequently, the weight of the finding on meat cooking is the highest; thus resulting in the most problematic area of food management practices for the respondents.

## 4.4  Information and Communication

This section presents the analysis of the four elements that were the focus of the information technology related questions presented to the research participants. The four elements are; information presentation format, perception on app usefulness, app's visual properties and information communication channels. These elements will be discussed in a logical sequence.

### 4.4.1  Information Presentation Format

Information presentation format determines the respondents' preferences on the format in which information on safe food management should be presented to them. This considers visual, textual, verbal or integrated modes of information. The findings reveal that the two most preferred information presentation formats are picture and text based information formats with Bayesian averages of 4.1 and 3.8 respectively, while audio-based information is least preferred. This finding on the preference of pictorial and textual information format (post-purchase) is consistent with Blanco et al. (2010) findings on pre-purchase consumer behaviours as they argued that pictorial information makes it easier for consumers to remember while textual information improves perception of information quality. However, the finding on the audio/verbal information format is not consistent with Kim and Lennon (2008) as they argued that verbal information is more "superior" than visual/pictorial information. Perhaps, this is based on the fact that their study was based on pre-purchase consumer behaviour, while this focuses on post-purchase.

Furthermore, participants were asked to provide insights into their thoughts regarding whether a safe food management app will be useful to them or not. This led to the term "perception on app usefulness". The findings indicate that about 21% of the respondents are confident of the fact that a mobile safe food management app will indeed be useful to them as individuals. From those who think a mobile safe food management would be useful to them, further questions were asked to investigate their preferences to provide insight into the style of design.

### 4.4.2  App's Visual Properties

App's visual properties provide insight into the visual requirements of the app for the respondent. The findings indicate that the two most relevant guideline providing insights into the preferred visual properties of the app on safe food management are the use of images such as Australian map or State-based maps and pie charts in order to illustrate information with Bayesian averages of 3.1 and 2.9 respectively. Whilst the least relevant guideline stated that "photographs should not be used and the images should be stylised", it can be inferred that pictures are again important to the consumers, when presenting information to them. This finding is in consonance with the finding in section 4.4.1 above. Other textual comments on the respondents' requirements of the application are as follows:

*"Bits of info could be in columns"; "Pictures paint 1000 words"; "cartoons are probably more amusing to the eye"; "Relevant laws and regulations with illustrations and practical examples with hazard and benefits"; "Simple to use"; and "I think it should be professional and creative like some options could be presented like a restaurant menu."*

### 4.5 Information Communication Channels

Information communication channel aims to determine the respondents' preferences on the mode of communication they would typically use when learning about the safe management of meat and other food products. The findings reveal that the three most important communication channels to the respondents are search engines such as Google, television and paper-based information channels such as newspapers, brochures, posters and pamphlets with Bayesian averages of 7.0, 6.6 and 6.4 respectively. Whilst the use of search engines was most preferred, it was however interesting to note that online videos such as YouTube, social networking sites, online groups or forums and online blogs were least preferred respectively. Perhaps, this is because 50% of the respondents were aged 50 years and above. However, it can be argued that this group of consumers merit particular attention as they spend more money on the average at food stores than their younger counterparts (Moschis et al. 2004) and that they have the poorest prognosis when food poisoning occurs (Johnson et al. 1998).

In summary, this section has presented the findings of the collected data, through an online survey, concerning the consumer focused aspect of the investigation. The analysis has provided insights into the areas of food handling practices that are most problematic for the respondents. The worst areas are kitchen hygiene practices, storage but more importantly cooking. In addition, we have generated insights on the factors influencing the respondents meat purchase in terms of information related attributes and food labels. It was revealed that the three most preferred information attributes are the use-by/best-before dates, the country of origin and the nutritional/health benefits respectively, while the three most preferred food labels are; meat standards Australia, national heart foundation and RSPCA approved farming. Furthermore, it was discovered that the respondents usually make use of search engines to find information when learning about safe management of meat. Graphical/ picture based information/ images were the most preferred mode of information and some respondents indeed believe that a mobile safe food management application would be useful to them as individuals.

## 5 DISCUSSION

Insights from the survey have contributed to on-going research with consumers into designing a food safety application. The survey has validated the existing literature and has confirmed aspects of Australian consumer preferences, issues and post-purchase food handling practices. It has also revealed consumer preferences on communication and information which may be useful for facilitating food related "good habit" *(see section 2)*. Having identified that there are challenges around safe meat cooking, there is more evidence that complements the existing literature on food poisoning and food wastage (DoH 2010; NSWFA 2014; Parfitt et al. 2010) which indicates that more effort should be placed on investigating the existing meat cooking apps. It is likely that testing existing applications will uncover how well it supports consumers to address – safe meat cooking, as well as other food management behaviours.

As earlier stated in the methodology section, the current focus for the next phase of this research involves the selection of three existing apps (text-based, graphics/picture-based and integrated) that most clearly addresses the safe food handling practice being targeted, in this case – safe meat cooking. Existing applications will be selected to facilitate:

- Easy and low cost investigation of three different information modalities which will support users (non-technical; food consumers) to become rapidly familiar with these modalities thereby enabling them to contribute to subsequent preferences for design by having provided them with some concrete examples to work from in the first instance (Houde and Hill 1997).

- A more effective collection of true human performance data as there is evidence to suggest that a high-fidelity prototype/fully functional prototype provides a more valid evaluation than a paper or low fidelity prototype (Lim et al. 2006). There is also a precedent for this type of approach in developing mobile solutions using existing applications Fleury et al. (2010).

Based on the aforementioned reasons, existing safe meat cooking apps have been selected. The three apps were selected based on the following criteria: it must be focused on end-consumers not professional food handlers; it must contain the appropriate meat cooking temperature and cooking time; it must be given a minimum of 4 out of 5 star rating. Due to copyright restrictions, the names of the apps cannot be specified in this paper. Following the selection of the three apps (text-based only, text & picture-based and integrated), heuristic evaluation based on Monkman and Kushniruk (2013) framework was conducted in order to identify problems with, arguably the three best existing apps on safe meat cooking, from an expert's perspective. The framework was selected due to its relevance for the context defined in this study. It is based on the following usability factors; screen, content, display, navigation and interactivity. The heuristics were developed (through the modification of design guidelines) for evaluating health app usability. Based on the frequency, impact and persistence of usability problems, a three-level severity scale was utilized with numbers 1, 2 and 3 representing mild, moderate and severe respectively. The Table 1 below shows the summary of heuristic evaluation of the three selected apps.

| Usability Factors | Heuristics | Evaluation Results (Text based app = T, Text and Pictures app = T & P while Integrated app = I) |
|---|---|---|
| Screens | Home screen | T & P was simple and engaging while T and I had problems with a severity score of 2 and 3 respectively. |
| Content | Hierarchy, promotion, positive tone, specific, colloquial, accurate, spacious – display content clearly on the page, personal and headings | All three were satisfactory with respect to hierarchy, specific, colloquial, accurate and the use of meaningful headings. However, they failed to specify the benefits of taking action. T had a severity rating of 3 each while T & P scored 2 each on "spacious" and "personal" heuristics respectively. |
| Display | Consistency, font, spacious – use white space and avoid clutter, location of content, images, contrast and accessibility | While all three apps were relatively satisfactory in terms of consistency, spacious, location of content and contrast. T had issues with font, with a severity score of 2; T could not be assessed for "images" as it is text based; and all three apps were not accessible to people with disabilities – scoring 3 on accessibility. |
| Navigation | Topics, orientation, back button, linear navigation, buttons, links and search | Whilst all three apps were relatively satisfactory on topics, back button, linear navigation and buttons; T & P did not enable easy access to home and menu screens thus scoring 2 on "orientation". T did not use links effectively neither did it include simple search options– scoring 3 each on "links" and "search". I also failed to include simple search and browse options, thus scoring 3. |
| Interactivity | Engage, print, multimedia and new media | All three apps did not provide tools to share content and feedback about experiences; thus scoring 3 on "engage". All three apps were not printer friendly; scoring 3 on "print". All three apps did not provide tools to explore new media such as Twitter or text messaging – scoring 3 on "new media". |

*Table 1: Summary of Heuristics Evaluation of the Three Apps*

The findings of the heuristics evaluation reveals that none of the apps currently comply with the best practice described by the framework as they were most deficient under "interactivity" as a usability factor. Therefore, the apps can proceed to the next stage of the consumer focused usability evaluation. This approach draws on Kushniruk (2002); Kushniruk et al. (1997); as usability testing is inserted into the design cycle, based on the evidence that it leads to 10-fold reduction in usability problems after implementation.

Using the apps as a high fidelity prototype in scenario-based focus group sessions, consumer focused usability evaluation will be conducted. This aims to identify the impact of the three information modalities on consumer understanding and to generate user requirements for a safe meat cooking app. The outcome of this phase provides richer and more in-depth insights into consumer requirements for a safe meat cooking app; thus complementing the design-related insights generated from the survey.

This leads us to identify the most effective mechanisms for delivering information to support and educate consumers on food safety with most particular focus on safe meat cooking practices.

# 6   CONCLUSION

This formative research is part of an investigation into the impact of a usability evaluation informed design approach on consumers' retention of knowledge on safe food handling practices. In this paper, we have presented findings from the nationwide survey which was conducted to identify problems with the current food handling practices of Australian consumers and their information and communication preferences. We have discovered that, of all domestic meat handling practices investigated, cooking is most susceptible to an unsafe outcome. Therefore, this has identified the main focus for the app as it leads to the next phase of the study.

Future work will involve the completion of the Phase 2 of the investigation through scenario-based focus group sessions with Australian consumers so as to identify the impact of the three information modalities on consumer understanding and to determine the best possible mechanism for presenting information to support education and hopefully behavioural change. Following this phase, usability evaluation of the developed app will be conducted through experimental trials using a control group and an experimental group *(see section 3)* to evaluate the impact of the design on the retention of knowledge on safe food handling practices over time. Therefore, it is anticipated that the outcome of this research moves one step closer towards influencing consumer safe food handling practices.

# 7   REFERENCES

ACCC. 2015. "Country of Origin." Australian Competition and Consumer Commission.

Aizaki, H., Sawada, M., and Sato, K. 2011. "Consumers' Attitudes toward Consumption of Cloned Beef. The Impact of Exposure to Technological Information About Animal Cloning," *Appetite* (57:2), pp. 459-466.

Al-Sakkaf, A. 2012. "Evaluation of Food Handling Practice among New Zealanders and Other Developed Countries as a Main Risk Factor for Campylobacteriosis Rate," *Food Control* (27:2), pp. 330-337.

Bamgboje-Ayodele, A., Ellis, L., and Turner, P. 2014. "Empowering Consumers in Food Value Chains: Preliminary Insights from an Investigation into the Development of Tools for Food Safety," *The 25th Australasian Conference on Information Systems (ACIS 2014)*, Auckland, New Zealand.

Bamgboje-Ayodele, A., Ellis, L., and Turner, P. 2015. "A Food Recall Case Study in Australia – Towards the Development of Food Safety Applications for Consumers," *International Journal of Food Studies* (in-press).

Bearth, A., Cousin, M.-E., and Siegrist, M. 2014. "Investigating Novice Cooks' Behaviour Change: Avoiding Cross-Contamination," *Food Control* (40:0), pp. 26-31.

Blanco, C. F., Sarasa, R. G., and Sanclemente, C. O. 2010. "Effects of Visual and Textual Information in Online Product Presentations: Looking for the Best Combination in Website Design," *European Journal of information systems* (19:6), pp. 668-686.

Bondarianzadeh, D., Yeatman, H., and Condon-Paoloni, D. 2011. "A Qualitative Study of the Australian Midwives' Approaches to Listeria Education as a Food-Related Risk During Pregnancy," *Midwifery* (27:2), pp. 221-228.

Charmaz, K. 2006. *Constructing Grounded Theory: A Practical Guide through Qualitative Analysis*. Pine Forge Press.

Chrysochou, P., Chryssochoidis, G., and Kehagia, O. 2009. "Traceability Information Carriers. The Technology Backgrounds and Consumers' Perceptions of the Technological Solutions," *Appetite* (53:3), pp. 322-331.

Cohen, S. 2003. "Maximum Difference Scaling: Improved Measures of Importance and Preference for Segmentation," *Sawtooth Software Conference Proceedings, Sawtooth Software, Inc*, pp. 61-74.

De Jonge, J., Van Trijp, H., Jan Renes, R., and Frewer, L. 2007. "Understanding Consumer Confidence in the Safety of Food: Its Two-Dimensional Structure and Determinants," *Risk analysis* (27:3), pp. 729-740.

DoH. 2010. "Food Poisoning and Contamination." Canberra, Australia: The Department of Health.

Erdem, S., Rigby, D., and Wossink, A. 2012. "Using Best–Worst Scaling to Explore Perceptions of Relative Responsibility for Ensuring Food Safety," *Food Policy* (37:6), pp. 661-670.

Ergönül, B. 2013. "Consumer Awareness and Perception to Food Safety: A Consumer Analysis," *Food Control* (32:2), pp. 461-471.

Finn, A., and Louviere, J. J. 1992. "Determining the Appropriate Response to Evidence of Public Concern: The Case of Food Safety," *Journal of Public Policy & Marketing* (11:2), pp. 12-25.

Fleury, A., Pedersen, J. S., and Larsen, L. B. 2010. "A Pragmatic Approach to Testing Issues in a Mobile Platform That Does Not yet Exist," *Adjunct Proceedings EuroITV 2010,* C. Peng, P. Vuorimaa, P. Näränen, C. Quico, G. Harboe and A. Lugmayr (eds.), Tampere University of Technology, Finland, pp. 262-263.

FSAI. 2003. "Consumer Attitudes to Food Safety in Ireland." Ireland: Food Safety Authority of Ireland.

Gilbert, S. E., Whyte, R., Bayne, G., Paulin, S. M., Lake, R. J., and van der Logt, P. 2007. "Survey of Domestic Food Handling Practices in New Zealand," *International journal of food microbiology* (117:3), pp. 306-311.

Grover, V. 1997. "A Tutorial on Survey Research: From Constructs to Theory."

Grunert, K. G., Hieke, S., and Wills, J. 2014. "Sustainability Labels on Food Products: Consumer Motivation, Understanding and Use," *Food Policy* (44:0), pp. 177-189.

Gustavsson, J., Cederberg, C., Sonesson, U., Van Otterdijk, R., and Meybeck, A. 2011. *Global Food Losses and Food Waste: Extent, Causes and Prevention*. FAO Rome.

Ha, Y., and Lennon, S. J. 2010. "Online Visual Merchandising (Vmd) Cues and Consumer Pleasure and Arousal: Purchasing Versus Browsing Situation," *Psychology & Marketing* (27:2), pp. 141-165.

Henderson, J., Coveney, J., Ward, P. R., and Taylor, A. W. 2011. "Farmers Are the Most Trusted Part of the Australian Food Chain: Results from a National Survey of Consumers," *Australian and New Zealand journal of public health* (35:4), pp. 319-324.

Houde, S., and Hill, C. 1997. "What Do Prototypes Prototype," *Handbook of human-computer interaction* (2), pp. 367-381.

Hyman, L., Lamb, J., and Bulmer, M. 2006. "The Use of Pre-Existing Survey Questions: Implications for Data Quality," *Proceedings of the European Conference on Quality in Survey Statistics*: Oxford University Press.

Jacob, C., Mathiasen, L., and Powell, D. 2010. "Designing Effective Messages for Microbial Food Safety Hazards," *Food Control* (21:1), pp. 1-6.

Johnson, A. E., Donkin, A., Morgan, K., Lilley, J. M., Neale, R. J., Page, R. M., and Silburn, R. 1998. "Food Safety Knowledge and Practice among Elderly People Living at Home," *Journal of Epidemiology and Community Health* (52:11), pp. 745-748.

Kim, M., and Lennon, S. 2008. "The Effects of Visual and Verbal Information on Attitudes and Purchase Intentions in Internet Shopping," *Psychology & Marketing* (25:2), pp. 146-178.

Kushniruk, A. 2002. "Evaluation in the Design of Health Information Systems: Application of Approaches Emerging from Usability Engineering," *Computers in biology and medicine* (32:3), pp. 141-149.

Kushniruk, A. W., Patel, V. L., and Cimino, J. J. 1997. "Usability Testing in Medical Informatics: Cognitive Approaches to Evaluation of Information Systems and User Interfaces," *Proceedings of the AMIA annual fall symposium*: American Medical Informatics Association, p. 218.

Kuttschreuter, M., Rutsaert, P., Hilverda, F., Regan, Á., Barnett, J., and Verbeke, W. 2014. "Seeking Information About Food-Related Risks: The Contribution of Social Media," *Food Quality and Preference* (37), pp. 10-18.

Lilavanichakul, A., and Boecker, A. 2013. "Consumer Acceptance of a New Traceability Technology: A Discrete Choice Application to Ontario Ginseng," *International Food and Agribusiness Management Review* (16:4), pp. 25-49.

Lim, Y.-k., Pangam, A., Periyasami, S., and Aneja, S. 2006. "Comparative Analysis of High-and Low-Fidelity Prototypes for More Valid Usability Evaluations of Mobile Devices," *Proceedings of the 4th Nordic conference on Human-computer interaction: changing roles*: ACM, pp. 291-300.

Lin, T. M., Lu, K.-Y., and Wu, J.-J. 2012. "The Effects of Visual Information in Ewom Communication," *Journal of Research in Interactive Marketing* (6:1), pp. 7-26.

Losasso, C., Cibin, V., Cappa, V., Roccato, A., Vanzo, A., Andrighetto, I., and Ricci, A. 2012. "Food Safety and Nutrition: Improving Consumer Behaviour," *Food Control* (26:2), pp. 252-258.

Louviere, J. J., and Flynn, T. N. 2010. "Using Best-Worst Scaling Choice Experiments to Measure Public Perceptions and Preferences for Healthcare Reform in Australia," *The Patient: Patient-Centered Outcomes Research* (3:4), pp. 275-283.

Louviere, J. J., and Woodworth, G. 1991. "Best-Worst Scaling: A Model for the Largest Difference Judgments." University of Alberta: Working Paper.

Lusk, J. L., and Briggeman, B. C. 2009. "Food Values," *American Journal of Agricultural Economics* (91:1), pp. 184-196.

Lyndhurst, B. 2008. "Research into Consumer Behaviour in Relation to Food Dates and Portion Sizes ", WRAP, United Kingdom.

Marley, A. 2009. "The Best-Worst Method for the Study of Preferences: Theory and Application." Psychology Press.

Mateus, T., Maia, R. L., and Teixeira, P. 2014. "Awareness of Listeriosis among Portuguese Pregnant Women," *Food Control* (46:0), pp. 513-519.

Mirowski, L., Marquez, L., Tamplin, M., and Turner, P. 2014. "Food Stability, Sensors and Value Chains: Issues and Challenges in Meat Traceability," in: *The Australian Research Council: Pathways to Market project*. Australia: University of Tasmania, pp. 1-9.

Monier, V. 2011. *Preparatory Study on Food Waste across Eu 27*. European Commission.

Monkman, H., and Kushniruk, A. 2013. "A Health Literacy and Usability Heuristic Evaluation of a Mobile Consumer Health Application," *MedInfo*, pp. 724-728.

Moschis, G., Curasi, C., and Bellenger, D. 2004. "Patronage Motives of Mature Consumers in the Selection of Food and Grocery Stores," *Journal of Consumer Marketing* (21:2), pp. 123-133.

Mullan, B., Allom, V., Fayn, K., and Johnston, I. 2014. "Building Habit Strength: A Pilot Intervention Designed to Improve Food-Safety Behavior," *Food Research International* (66), pp. 274-278.

Mullan, B., and Wong, C. 2010. "Using the Theory of Planned Behaviour to Design a Food Hygiene Intervention," *Food Control* (21:11), pp. 1524-1529.

Nesbitt, A., Thomas, M. K., Marshall, B., Snedeker, K., Meleta, K., Watson, B., and Bienefeld, M. 2014. "Baseline for Consumer Food Safety Knowledge and Behaviour in Canada," *Food Control* (38:0), pp. 157-173.

NSWFA. 2014. "Food Poisoning." Australia: New South Wales Food Authority.

Ovca, A., and Jevšnik, M. 2009. "Maintaining a Cold Chain from Purchase to the Home and at Home: Consumer Opinions," *Food Control* (20:2), pp. 167-172.

OzFoodNet. 2012. "Monitoring the Incidence and Causes of Diseases Potentially Transmitted by Food in Australia: Annual Report of the Ozfoodnet Network, 2010," *Communicable diseases intelligence quarterly report* (36:3), pp. E213-241.

Parfitt, J., Barthel, M., and Macnaughton, S. 2010. "Food Waste within Food Supply Chains: Quantification and Potential for Change to 2050," *Philosophical transactions of the royal society B: biological sciences* (365:1554), pp. 3065-3081.

Phillip, S., and Anita, E. 2010. "Efficacy of the Theory of Planned Behaviour Model in Predicting Safe Food Handling Practices," *Food Control* (21:7), pp. 983-987.

Polkinghorne, R., Watson, R., Thompson, J., and Pethick, D. 2008. "Current Usage and Future Development of the Meat Standards Australia (Msa) Grading System," *Animal Production Science* (48:11), pp. 1459-1464.

Reid, M., Li, E., Bruwer, J., and Grunert, K. 2001. "Food-Related Lifestyles in a Cross-Cultural Context: Comparing Australia with Singapore, Britain, France and Denmark," *Journal of Food Products Marketing* (7:4), pp. 57-75.

Rimal, R. N. 2000. "Closing the Knowledge-Behavior Gap in Health Promotion: The Mediating Role of Self-Efficacy," *Health Communication* (12:3), pp. 219-237.

Rogers, Y., Sharp, H., and Preece, J. 2011. *Interaction Design: Beyond Human-Computer Interaction*. Chichester: John Wiley & Sons.

Rompotis, C. J., Grove, J. R., and Byrne, S. M. 2014. "Benefits of Habit-Based Informational Interventions: A Randomised Controlled Trial of Fruit and Vegetable Consumption," *Australian and New Zealand journal of public health* (38:3), pp. 247-252.

Sainsbury, K., Mullan, B., and Sharpe, L. 2013. "Gluten Free Diet Adherence in Coeliac Disease. The Role of Psychological Symptoms in Bridging the Intention–Behaviour Gap," *Appetite* (61:0), pp. 52-58.

Shim, S.-M., Seo, S. H., Lee, Y., Moon, G.-I., Kim, M.-S., and Park, J.-H. 2011. "Consumers' Knowledge and Safety Perceptions of Food Additives: Evaluation on the Effectiveness of Transmitting Information on Preservatives," *Food Control* (22:7), pp. 1054-1060.

Smith, D., and Riethmuller, P. 1999. "Consumer Concerns About Food Safety in Australia and Japan," *International Journal of Social Economics* (26:6), pp. 724-742.

Stenger, K. M., Ritter-Gooder, P. K., Perry, C., and Albrecht, J. A. 2014. "A Mixed Methods Study of Food Safety Knowledge, Practices and Beliefs in Hispanic Families with Young Children," *Appetite* (83:0), pp. 194-201.

Taché, J., and Carpentier, B. 2014. "Hygiene in the Home Kitchen: Changes in Behaviour and Impact of Key Microbiological Hazard Control Measures," *Food Control* (35:1), pp. 392-400.

Van Rijswijk, W., and Frewer, L. J. 2012. "Consumer Needs and Requirements for Food and Ingredient Traceability Information," *International Journal of Consumer Studies* (36:3), pp. 282-290.

van Teijlingen, E., and Hundley, V. 2001. "The Importance of Pilot Studies," *Social research update* (Winter 2001:35), pp. 1-4.

Voordouw, J., Cornelisse-Vermaat, J. R., Pfaff, S., Antonides, G., Niemietz, D., Linardakis, M., Kehagia, O., and Frewer, L. J. 2011. "Preferred Information Strategies for Food Allergic Consumers. A Study in Germany, Greece, and the Netherlands," *Food Quality and Preference* (22:4), pp. 384-390.

Watson, R., Gee, A., Polkinghorne, R., and Porter, M. 2008. "Consumer Assessment of Eating Quality– Development of Protocols for Meat Standards Australia (Msa) Testing," *Animal Production Science* (48:11), pp. 1360-1367.

Williams, P., Stirling, E., and Keynes, N. 2004. "Food Fears: A National Survey on the Attitudes of Australian Adults About the Safety and Quality of Food," *Asia Pacific Journal of Clinical Nutrition* (13:1), pp. 32-39.

Wynd, C. A., Schmidt, B., and Schaefer, M. A. 2003. "Two Quantitative Approaches for Estimating Content Validity," *Western Journal of Nursing Research* (25:5), pp. 508-518.